\documentclass[preprint,showpacs,preprintnumbers,amsmath,amssymb,prl,aps]{revtex4-1}
\usepackage{graphicx}
\usepackage{bm}
\usepackage{float} 
\usepackage{amsmath}
\usepackage{gensymb}
\usepackage{epstopdf}
\usepackage[usenames, dvipsnames]{color}

\begin{document}

\title{Emergent electronic structure of CaFe$_2$As$_2$}

\author{Khadiza Ali and Kalobaran Maiti}

\altaffiliation{Corresponding author: kbmaiti@tifr.res.in}
\affiliation{Department of Condensed Matter Physics and
Materials' Science, Tata Institute of Fundamental Research, Homi
Bhabha Road, Colaba, Mumbai - 400 005, INDIA.}

\date{\today}

\begin{abstract}
{\bf CaFe$_2$As$_2$ exhibits collapsed tetragonal (cT) structure and varied exotic behavior under pressure at low temperatures that led to debate on linking the structural changes to its exceptional electronic properties like superconductivity, magnetism, etc. Here, we investigate the electronic structure of CaFe$_2$As$_2$ forming in different structures employing density functional theory. The results indicate better stability of the cT phase with enhancement in hybridization induced effects and shift of the energy bands towards lower energies. The Fermi surface centered around $\Gamma$ point gradually vanishes with the increase in pressure. Consequently, the nesting between the hole and electron Fermi surfaces associated to the spin density wave state disappears indicating a pathway to achieve the proximity to quantum fluctuations. The magnetic moment at the Fe sites diminishes in the cT phase consistent with the magnetic susceptibility results. Notably, the hybridization of Ca 4$s$ states (Ca-layer may be treated as a charge reservoir layer akin to those in cuprate superconductors) is significantly enhanced in the cT phase revealing its relevance in its interesting electronic properties.}
\end{abstract}


\maketitle

\section{Introduction}

\begin{figure}[H]
 \centering
 \vspace{-2ex}
 \includegraphics[scale=0.4]{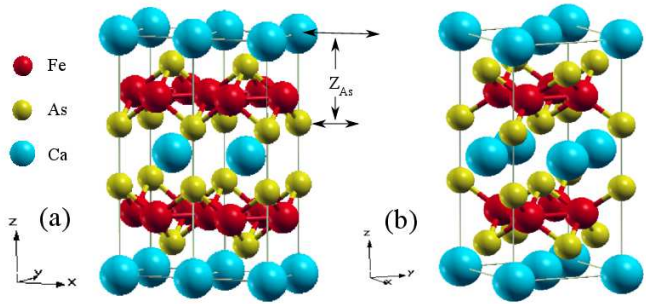}
 \caption{Crystal structure of CaFe$_2$As$_2$. (a) Double unit cell of the tetragonal structure and (b) single unit cell of the orthorhombic structure.}
 \label{Fig1_struct}
 \vspace{-2ex}
\end{figure}

CaFe$_2$As$_2$ exhibits a rich temperature - pressure phase diagram and provides an ideal platform to study the interplay between magnetism and superconductivity in Fe-based compounds, where the superconductivity is believed to be driven by magnetic fluctuations. The crystal structure of CaFe$_2$As$_2$ at room temperature and ambient pressure is tetragonal possessing $I4/mmm$ space group with lattice parameters $a$ = 3.8915(2)\AA, $c$ = 11.690(1)\AA, and $z_{As}$ = 0.372(1)\cite{phasetransition_NiN_PRB08,structure_Tompsett_physicab2009,neutron-kreyssing-PRB08}. Upon cooling at ambient pressure, it undergoes a concomitant transition to spin density wave (SDW) state and orthorhombic (O) structure from its room temperature paramagnetic tetragonal (T) structure at about 170 K \cite{phasetransition_NiN_PRB08}. The crystal structures in ambient conditions are shown in Fig. \ref{Fig1_struct}. On application of pressure ($P$), {\it both} the structural and magnetic transitions get suppressed, and a collapsed tetragonal (cT) phase emerges with no magnetic long range order \cite{neutron-kreyssing-PRB08,ct-pratt-goldman-PRB09,ct-sc-MKumar-PRB08,neutron-Goldman-PRB08,sc-ct-torikachi-PRL08}. Superconductivity can be achieved in this material at an intermediate pressure range or via suitable electron or hole doping \cite{sc-ct-torikachi-PRL08,sc_cafe2as2_saha_PRB_2012}.

\subsection*{Historical Background}

Electronic structure calculations based on first principle approaches were employed extensively
to address the ground state properties, Fermi surface and pressure induced effects \cite{structure_Tompsett_physicab2009,dft_yildrim_prl2009,ganesh-EuFe2As2, dft_Colonna_prb_2011,dft_Zhang_prb_2009}. While ligand $p$ states play major role in the electronic properties of cuprate superconductors  (the most studied unconventional superconductors), the main contribution at the Fermi level, $\epsilon_F$ in the Fe-based materials arises from the Fe 3$d$ orbitals. These systems are more complex than cuprates due to the contribution of multiple bands at $\epsilon_F$ and drastic changes in the Fermi surface topology arising from magnetic and structural transitions. For example, the two dimensional hole pockets at $\Gamma$-point in the tetragonal phase exhibit a transition to three dimensional topology at low temperatures \cite{ARPES-Liu,ganesh-CaFe2As2} and in cT phase, they disappear due to the shift of the energy bands. The disappearance of the Fermi surface in cT phase rules out the possibility of Fermi surface nesting necessary for the transition to the SDW state. The Fe-based systems, in general, are found to be moderately correlated and inclusion of electron correlation strength, $U$ in the calculations renormalizes the energy bands without drastic change in the ground state properties \cite{dft_Diehl_prb_2014,dft_Kurmaev_prb_2009,ganesh-FeTeSe,silke-arxiv,silke-PRB}. Jean Diehl {\it et al.}\cite{dft_Diehl_prb_2014} showed that the Hund's coupling, $J$ ($\sim$0.8 eV) influences the electronic structure more strongly than the changes found due to $U$ ($\sim$ 4 eV). They also showed that upon transition from T to cT phase, the $d_{xy}$ bands become the least renormalized orbital from the highest renormalized one, which was attributed to increased hybridization of Fe 3$d_{xy}$ orbital with As 4$p_x$ and 4$p_y$ orbitals. The appearance of the cT phase under pressure and/or on suitable doping \cite{dft-Sanna-PRB12,sc_cafe2as2_saha_PRB_2012} has been attributed to the increased hybridization among As 4$p$ orbitals along the compressed $c$ axis. Yildrim {\it et al.} \cite{dft_yildrim_prl2009} found that the Fe spin state strongly depends on the $c$ axis reduction in cT phase. Based on fixed moment calculation, he showed that reduction by about half of the Fe magnetic moment corresponds to the $c$ axis collapse to the experimental value. Thus, there is a finite magnetic moment present at the Fe sites in the cT phase when it is not superconducting. Evidently, understanding of the importance of Fe moment in superconductivity is still at its infancy and needs further investigation. These results indicate that the electronic structure is very sensitive to the distance of As sites from the Fe plane, which is termed as {\it `pnictogen height'}. In the antiferromagnetic (AFM) orthorhombic (O) phase, the spins show antiparallel coupling along the longer axis, $a$ and parallel coupling along shorter axis, $b$. This has been theoretically found by many groups \cite{dft_yildrim_prl2009,dft_Deepa_njp_2009} but a satisfactory resolution of the preferred spin coupling along $a$ and $b$ axis is not yet found. First principle calculations under hydrostatic\cite{dft_Zhang_prb_2009,dft_Widom_PRB_2013} and uniaxial\cite{dft_Tomi_prb_2012} pressure conditions reproduced the first order transition to cT phase from AFM O/T phase accompanied by quenching of Fe magnetic moments and also rule out the possibility of an intermediate tetragonal phase in CaFe$_2$As$_2$ in contrary to various experiments \cite{neutron-kreyssing-PRB08,neutron_goldman_prb,sc-p-park-JPCM08}, which is different from the behavior in BaFe$_2$As$_2$ at intermediate pressure range \cite{dft_Tomi_prb_2012}.

\begin{figure}[H]
 \centering
 \vspace{-2ex}
 \hspace{0ex}
 \includegraphics[scale=0.4]{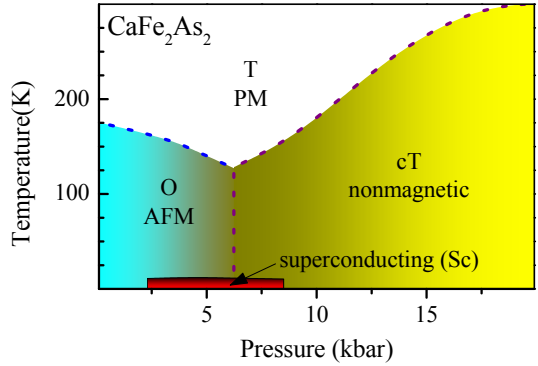}
 \vspace{-28ex}
 \caption{(a) Temperature - Pressure phase diagram of CaFe$_2$As$_2$ \cite{sc-ct-torikachi-PRL08}. The dotted line represents notional phase boundary between various structural and magnetic phases.}
 \label{Fig2_schem}
 \vspace{-2ex}
\end{figure}

In the experimental front, the pressure induced effects in AFe$_2$As$_2$ (A = Ca, Ba, Sr) has been studied by many groups resulting into conflicting conclusions as discussed below. Superconductivity in BaFe$_2$As$_2$ \& SrFe$_2$As$_2$ containing alkaline earth elements larger than Ca was found at a reasonably high pressure ($P$ $\approx$ 4 GPa and 3.2 GPa, respectively\cite{Alireza-JPCM09}) with $T_c$ of about 29 K and 27 K \cite{sc_p_Awadhesh_Mani_epl_2009,sc_p_Hisashi_Kotegawa_JPSJ_2009, sc_p_Colombier_prb_2009,34K_sc_p_Hisashi_Kotegawa_JPSG_2009,sc_p_Kazumi_Igawa_JPSG_2009,ct-sc-MKumar-PRB08}. On the other hand, CaFe$_2$As$_2$ shows superconductivity with $T_c$ upto 12 K in a low pressure range from  0.2 to 0.9 GPa \cite{sc-ct-torikachi-PRL08} although the smaller atomic size of Ca is expected to employ less strain into the system. A typical temperature-pressure phase diagram shown in Fig. \ref{Fig2_schem} exhibits the experimental results obtained by Torikachvili {\it et al.} \cite{sc-ct-torikachi-PRL08}, which is consistent with the results in other studies \cite{sc_p_Torikachvili_prb_2009}. In Fig. \ref{Fig2_schem}, we observe that the transition to orthorhombic AFM phase is suppressed upon increasing pressure and disappears above $P$ = 0.35 GPa with the emergence of the cT phase. The transition temperature from T to cT phase gradually increases with pressure. The superconductivity appears at an intermediate $P$ range from 0.3 GPa to 0.9 GPa. In fact, superconductivity with $T_c$ $\approx$ 10 K at 0.69 GPa pressure has been detected by several groups \cite{sc-p-park-JPCM08,sc_p_lee_prb_2009}. Subsequent experiments, however, did not find superconductivity although cT phase could be reached on application of pressure. It was suggested that the earlier measurements carried out using convectional liquid medium clamp pressure cell might have made the pressure quasi hydrostatic due to freezing of the medium. These later results obtained using helium as a pressure medium question the relation of superconductivity to cT phase and in line with other views suggesting non-magnetic behavior in cT phase that rules out spin-fluctuation. Evidently, the observance/non-observance of superconductivity under pressure, non-observance of other structural phases under pressure, no superconductivity in ambient conditions, {\it etc.} are curious and remains to be an outstanding puzzle in this field.

\subsection*{Objective}

A study of the reported resistivity data exhibiting superconductivity reveal an interesting scenario; the transitions in the resistivity data at $P$ = 0.23 GPa are significantly broad. On the other hand, the results exhibiting absence of superconductivity show sharp transitions throughout the pressure range involving the transitions from T to O, O to cT and T to cT phases. Moreover, the superconductivity always found to appear at a pressure where the system is in the proximity of the transition to cT phase or some volume fraction is already in cT phase. Thus, the origin of superconductivity in this material appears to have relation to the multi-crystallographic phases within the intermediate pressure regime. Nuclear magnetic resonance \cite{NMR_baek_prb,NMR_kawasaki_iop}, $\mu$ spin rotation \cite{msr_goko_prb}, and neutron diffraction studies \cite{neutron-kreyssing-PRB08,neutron_goldman_prb,sc_cafe2as2_Proke_PRB_2010} support the possibilities of coexisting O, cT and T phases. Some studies indicated signature of an unidentified third phase appearing in the intermediate pressure regime \cite{sc_p_lee_prb_2009}. While the magnetic order in single or multiple phase scenario can be attributed to the O phase, it is difficult to delineate the phase responsible for superconductivity. Clearly, superconductivity does not appear to be related to pure cT or O phase under pressure \cite{ct_sc_absence_Yu_PRB_2009}. Then the question is; how the superconductivity sets in CaFe$_2$As$_2$ under pressure? How non hydrostatic pressure helps to get supeconductivity while hydrostatic pressure does not?  Superconductivity also appears via doping of charge carriers in these material - how different is that mechanism from the pressure induced one?

Here, we studied the detailed electronic structure of CaFe$_2$As$_2$ employing density functional theoretical (DFT) methods at varied conditions with an emphasis on the tendency of the system to evolve to a certain electronic structure due to change in structural parameters induced by the pressure. We observe interesting spectral weight transfer with the application of pressure leading to complex change in Fermiology of the system. The covalency between Ca, As and Fe states appear to play the key role in deriving the electronic properties of this system.

\section{Results}

The electronic structure of a material is very sensitive to the change in crystal structure as the hybridization of various electronic states depends on the bond length, distortion in the lattice structure (leading to a change in bond angle) etc. Often it is found that covalency induces key interactions leading to distortion in the crystal structure \cite{Andersen-PRL,CaSrRuO-PRB,BaIrO3-PRB} that derives the ground state properties.

\subsection*{Density of States}

\begin{figure}
 \centering
 \vspace{-2ex}
 \includegraphics[scale=.5]{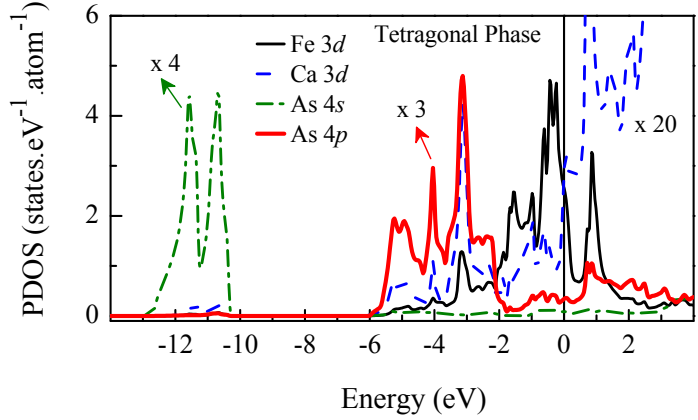}
 \vspace{-36ex}
 \caption{Calculated partial density of states of Ca, Fe and As atoms in the tetragonal phase (total contributions). The PDOS of As is multiplied by 3 and Ca PDOS by 20 to show their contribution clearly. Here, `0' in energy scale denotes the Fermi level, $\epsilon_F$.}
 \label{fig:tdos}
 \vspace{-2ex}
\end{figure}

In Fig. \ref{fig:tdos}, we show the calculated density of states (DOS) of CaFe$_2$As$_2$ in tetragonal phase. There are two distinctly separated energy regions in the figure contributed by As 4$s$ shallow core levels (-14 to -10 eV) and the valence electronic states above -6 eV. In the valence band regime, we observe signatures of hybridization between Fe 3$d$ and As 4$p$ states. Major contribution from As 4$p$ states appear in the bonding energy bands between -6 to -2 eV energies. The contribution for antibonding states appear above -2 eV and possess dominant Fe 3$d$ character. The valence electrons in the vicinity of the Fermi level, $\epsilon_F$ consist of essentially Fe 3$d$ states. The Fe-As hybridization leads to finite contribution from As 4$p$ states ($\sim$~4\% of the total contribution at $\epsilon_F$). Interestingly, the contribution of the Ca 3$d$ states are also found to be significant ($\sim$~2.4\%). This suggests that although Fe 3$d$ states play the dominant role in the electronic properties of CaFe$_2$As$_2$ as also found in earlier studies, the role of As 4$p$ and Ca 3$d$ states are non-zero due to covalency between these states with Fe 3$d$ states. The contribution from Ca 4$s$ states, however, is almost negligible ($\sim$~0.2\%) in the vicinity of $\epsilon_F$.

\begin{figure}
 \centering
 \vspace{-2ex}
 \includegraphics[scale=0.5]{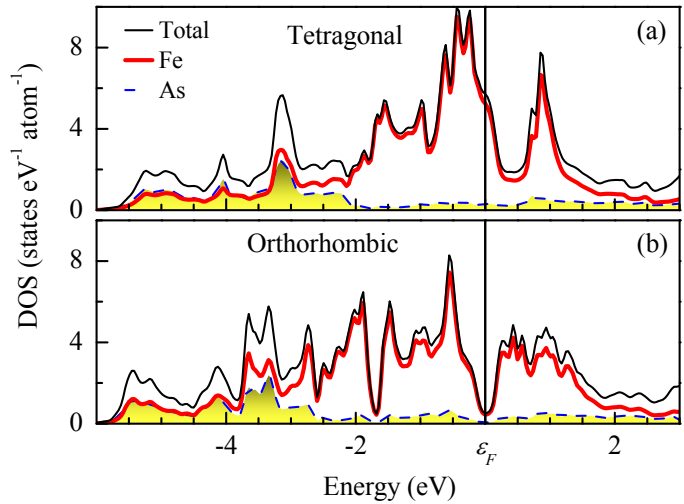}
 \vspace{-30 ex}
 \caption{Calculated total density of states and the contributions from Fe \& As in the (a) tetragonal and (b) orthorhombic phase.}
 \label{Fig4-STRcomp}
 \vspace{-2ex}
 \end{figure}

 At ambient pressure, CaFe$_2$As$_2$ undergoes a magneto-structural transition from paramagnetic tetragonal to antiferromagnetic orthorhombic phase at 170 K. In Fig. \ref{Fig4-STRcomp}, we show the calculated DOS for tetragonal and orthorhombic structures to investigate the effect in the electronic structure due to the above mentioned transitions. While the DOS at $\epsilon_F$ in the tetragonal structure is large and has a decreasing slope with the increase in energy, a {\it pseudogap} like feature appears in the magnetically ordered phase as shown in Fig. \ref{Fig4-STRcomp}(b). $\epsilon_F$ is situated at the middle of the {\it pseudogap} exhibiting particle-hole symmetry in its close proximity. The position and width of the pseudogap observed here matches well with the LDA+DMFT results at ambient pressure\cite{DMFT_Yin_naturematerials_2011} (here, LDA stands for local density approximation and DMFT for dynamical mean field theory). While there is no apparent change in overall Fe 3$d$ bandwidth due to the transitions, significant change is observed in the details, such as the shift of the Fe 3$d$ spectral weight away from $\epsilon_F$ leading to significant broadening of the sharp feature above $\epsilon_F$. A new feature appear around -2 eV energy. The center of mass of the As 4$p$ contributions shifts to lower energy along with an enhanced contribution near $\epsilon_F$ indicating a stronger mixing of the Fe 3$d$ and As 4$p$ electronic states. Such enhanced mixing/covalency between Fe 3$d$ and As 4$p$ states might be the driving force for the magnetic and structural transitions in the ground state.

\begin{figure}
 \centering
 \vspace{-2ex}
 \includegraphics[scale=0.5]{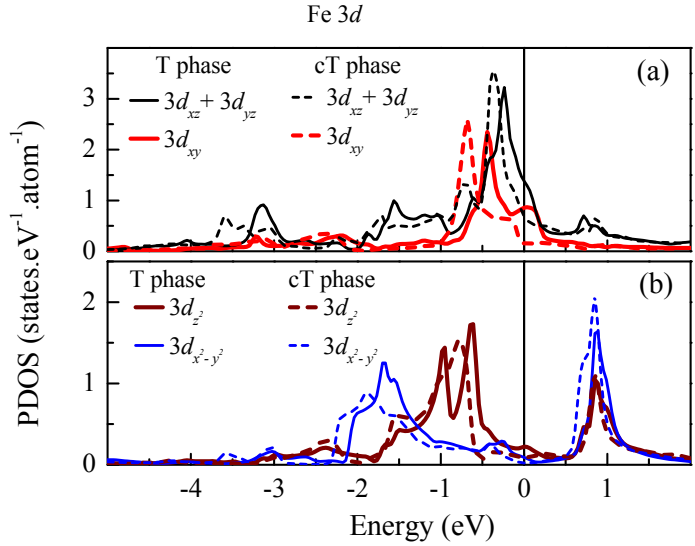}
  \vspace{-28 ex}
 \caption{Comparison of the partial density of states of various Fe 3$d$ orbitals in tetragonal (T)-phase (solid lines) and collapsed tetragonal (cT) phase (dashed lines). (a) 3$d_{xz}$+3$d_{yz}$ and 3$d_{xy}$ are shown by thin black line and thick red line, respectively. (b) 3$d_{z^2}$ and 3$d_{x^2-y^2}$ states are shown by thick maroon line and thin blue line, respectively.}
 \label{Fig5TcT-Fe3d}
 \vspace{-2ex}
\end{figure}

In Fig. \ref{Fig5TcT-Fe3d}, we superimpose the Fe 3$d$ partial density of states (PDOS) for collapsed tetragonal and tetragonal phases to have a comparative study. Fe atoms in tetragonal structure experience tetrahedral crystal field created by the As atoms. A regular tetrahedral crystal field splits the Fe 3$d$ levels into a doubly degenerate $e_g$ band and a triply degenerate $t_{2g}$ band. The distortions in the crystal structure of CaFe$_2$As$_2$ leads to a deviation from the regular tetrahedral field and the degeneracy of the crystal field split bands are lifted with $d_{xy}$ band slightly shifted above the doubly degenerate $d_{xz}$ \& $d_{yz}$ bands. The contributing energy range and width derived by the hopping interaction terms remain very similar for all these bands as manifested in the figure. The $t_{2g}$ band consisting of $d_{xz}$, $d_{yz}$ and $d_{xy}$ orbitals lies close to $\epsilon_F$ and are the major contributors to the Fermi surface. The orbitals $d_{x^2-y^2}$ and $d_{z^2}$ possessing $e_g$ symmetry are situated far below the Fermi level.

In the cT phase, the As-Fe-As angle (= 106.13$\degree$) deviates from the regular tetrahedral angle at ambient condition ($\approx$ $109.47\degree$) and gives rise to additional distortions in the FeAs$_4$ tetrahedra that enhances further the crystal filed splitting \cite{tetrahedra_screp_sarapov2014}. In Fig. \ref{Fig5TcT-Fe3d}(a) and (b), it appears that all the energy bands in cT phase represented by dashed lines are shifted towards lower energies relative to those in T phase. The energy bandwidth of $d_{xz}$ + $d_{yz}$ bands increases slightly in cT phase compared to that in T phase while the other $t_{2g}$ and $e_g$ bandwidths remain unchanged. This implies that in cT phase, the hybridization between Fe-As along $z$ direction enhances as expected due to the compression along $c$-axis; the in-plane orbitals remain almost unaffected. In T phase, major contribution at $\epsilon_F$ comes from ($d_{xz}$ + $d_{yz}$) and $d_{xy}$ electronic states. The $d_{xy}$ band shifts below $\epsilon_F$ in cT phase and thus, contribution at $\epsilon_F$ in cT phase arises primarily from degenerate ($d_{xz}$ + $d_{yz}$) orbitals. Evidently, the carrier density at $\epsilon_F$ is significantly less in the cT phase due to the shift of the energy bands, which is in contrast to the transport data indicating a decrease in electrical resistivity in the cT phase compared to the tetragonal phase \cite{sc-ct-torikachi-PRL08}. While more studies are required to understand this phenomena, the decrease in resistivity may be attributed to the enhancement of the mobility of charge carriers; the increase in hybridization of Fe ($d_{xz}$ + $d_{yz}$) and As 4$p$ states facilitates higher degree of the hopping of charge carriers.

\begin{figure}
\centering
\vspace{-8ex}
\includegraphics[trim={0cm 0cm 0cm 0cm}, clip, scale=0.5]{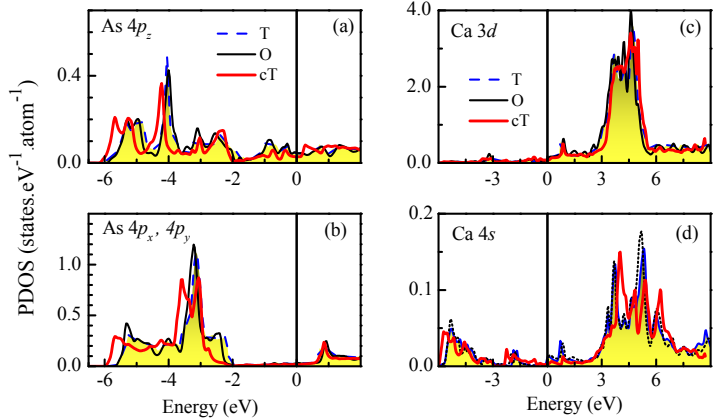}
\vspace{-36 ex}
\caption{Partial Density of states of (a) As 4$p_z$, (b) As 4$p_x$ + 4$p_y$, (c) Ca 3$d$ and (d) Ca 4$s$ in collapsed tetragonal (red thick solid line), tetragonal (blue dashed line) and orthorhombic (black thin solid line) phases.}
\label{Fig6TcT-As4p}
\vspace{-2ex}
\end{figure}

The compression along $c$ axis in the cT phase influences the hybridization among As 4$p$ states too \cite{dft_yildrim_prl2009} leading to notable changes in their density of states. In Fig. \ref{Fig6TcT-As4p}(a) and (b), we show the PDOS corresponding to As 4$p_z$ and the degenerate As (4$p_x$+4$p_y$) energy bands. The results corresponding to the orthorhombic phase shown in the figure were calculated without considering magnetic order to study the effect on electronic structure due to the change in structural parameters only. From the figure, it is clear that all the As 4$p$ bands in O and T phases are essentially identical while the $p$ bands in cT phase exhibit a shift towards lower energies relative to those in T and O phases - a trend similar to the findings in Fe 3$d$ cases. In addition, there are several subtle changes observed for various $p$ orbitals. For example, the overall width of the As 4$p_z$ is somewhat larger in the cT phase compared to T phase due to enhancement of DOS near the band edges suggesting an effective increase in hybridization among the $p_z$ orbitals. The energy distribution of the density of states corresponding to the $p_x$ and $p_y$ electronic states exhibit spectral weight shift near the top of the As 4$p$ bands (see around -2 eV in the figure). The unoccupied part appears to be close to each other in all the three cases studied. The changes in the bonding bands between -6 to -2 eV indicate signature of enhancement of covalency in the system in the cT phase.

The results for the Ca 3$d$ and 4$s$ PDOS is shown in the Fig. \ref{Fig6TcT-As4p}(c) and (d), respectively. Ca 3$d$ states contribute primarily above $\epsilon_F$ (energy range 3 to 6 eV) with small contributions at and below $\epsilon_F$. The intensity of Ca 4$s$ states is distributed over larger energy range. While the major contribution of 4$s$ states appear between 3 to 8 eV above $\epsilon_F$, there is finite contribution below -3 eV energy. Interestingly, the contributions below $\epsilon_F$ exhibit a shift towards lower energies in the cT phase as found in Fe and As cases keeping the PDOS above $\epsilon_F$ within almost the same energy range in all the structural phases. In the crystal structure, the Ca layers are sandwiched between two As layers and the changes in Ca states below $\epsilon_F$ can be attributed to the hybridization of Ca 4$s$ states with the electronic states associated to As in the valence band. The compression \cite{ct_vanroekeghem_prb_2016} along $c$-axis due to the transition to the cT phase reduces the separation between the Ca and As atomic layers and the hybridization between Ca 4$s$ and As 4$p$ will be enhanced. It is clear that although the Ca contribution in the valence band regime is small, it is difficult to ignore the role of Ca in the structural changes and their electronic structure as well \cite{casrruo-core-PRB07,casrruo-core-SL14}. Almost all the high temperature (unconventional) superconductors exhibit effective two dimensional electronic structure, where the conduction sheet is separated by an insulating layer, often called {\it 'charge reservoir'} layer. It is believed that the cations with $s$ electrons in the charge reservoir layer plays an important role in the pairing interactions, renormalizing disorder induced effects, etc \cite{charge-reservoir,charge-reservoir_book}. In the cuprate superconductors, elements like La, Hg, Pb, Bi form the charge reservoir layers and the Cu-O layers form the two dimensional conduction layers. In the present case, Ca layer sandwiched between conducting FeAs layers appears to play similar role and hence enhanced interaction of Ca 4$s$ states in the cT phase appears to have great implication for the onset of superconductivity in these systems.

\begin{figure}
\centering
\includegraphics[natwidth=620,natheight=820,width=0.6\textwidth]{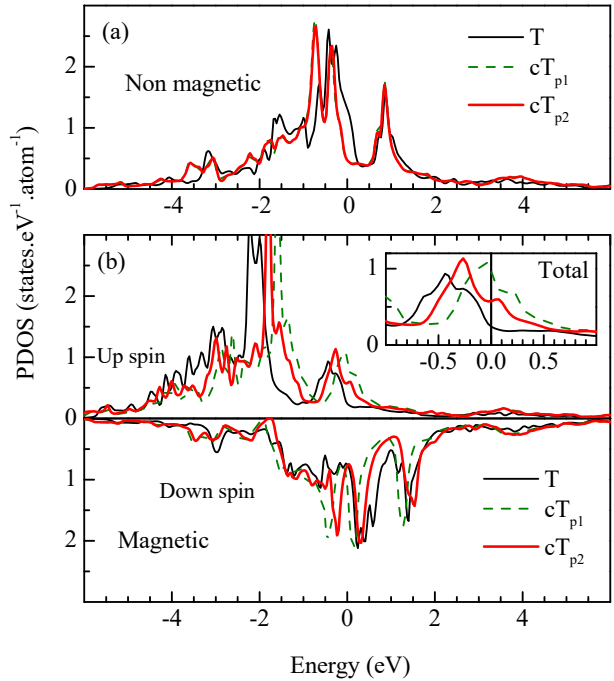}
 \vspace{-12 ex}
 \caption{Fe 3$d$ partial density of states in tetragonal (T; thin solid line) and collapsed tetragonal structures at pressures, $P$ = 0.47 GPa (cT$_{p1}$; dashed green line) and $P$ = 0.63 GPa (cT$_{p2}$; thick solid line) in (a) non-magnetic and (b) magnetically ordered phases. The inset in (b) shows the total Fe 3$d$ PDOS at different structures. The definition of lines is kept similar to the main part of the figure.}
\label{Fig7-pres}
\end{figure}
The results for cT phase discussed so far correspond to the lattice constants found experimentally for the pressure of 0.63 GPa. Now, we study the density of states at different pressures utilizing the structural parameters obtained experimentally. The DOS calculated for tetragonal, collapsed tetragonal phases at $P$ = 0.47 GPa (cT$_{p1}$) and $P$ = 0.63 GPa (cT$_{p2}$) are superimposed in Fig. \ref{Fig7-pres}. The Fe 3$d$ PDOS in the non-magnetic phase shown in Fig. \ref{Fig7-pres}(a) exhibit significant intensity at $\epsilon_F$ in the tetragonal phase. With the application of pressure, Fe 3$d$ PDOS shift to lower energies reducing the intensity at $\epsilon_F$ to almost zero. Interestingly, the results for cT$_{p1}$ and cT$_{p2}$ appear identical indicating less dependence of the non magnetic phase on pressure once the {\it pseudogap} like feature at $\epsilon_F$ is present in the electronic structure. The overall bandwidth appears to be similar in every case.

The magnetic solution shown in Fig.  \ref{Fig7-pres}(b) exhibits large separation between the up and down spin bands in the tetragonal phase with an exchange splitting of the order of 2 eV. With the application of pressure, up and down spin Fe 3$d$ PDOS shift towards each other reducing the exchange splitting with the increase in pressure. This change, however, is not monotonic; the data corresponding to 0.47 GPa shows much larger shift and smaller exchange splitting compared the case for $P$ = 0.63 GPa. Since the bottom of the valence band remains almost same in every case, the change at the top of the band also leads to an increase in bandwidth. Thus, the width of the valence band for $P$ = 0.47 GPa appears to be the largest. The transition to the magnetically ordered phase in tetragonal structure leads to a large dip ($pseudogap$) in DOS at $\epsilon_F$. Interestingly, the dip in the magnetically ordered phase reduces with the application of pressure in cT phase as shown in the inset of Fig. \ref{Fig7-pres}(b), where sum of the up and down spin Fe 3$d$ contributions are plotted. This indicates poorer degree of stabilization of the magnetic phase in cT structures and the electronic properties at $P$ = 0.47 GPa is closer to the proximity of instability in magnetic order relative to the other cases.

\begin{figure}
 \centering
 \vspace{-2ex}
 \includegraphics[trim={0cm 0cm 0cm 0cm}, clip, scale=.5]{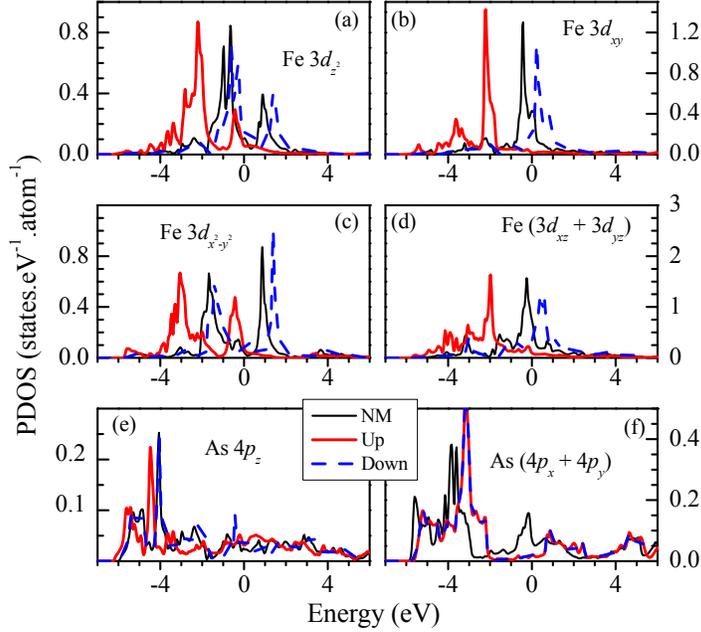}
\vspace{-20 ex}
 \caption{Comparison of the partial Density of states of different $d$ orbitals calculated for magnetic and non Magnetic (NM) configurations for tetragonal structure. The details of the $d$ symmetries are defined as legends in each block of the plots. NM PDOS, and Up \& Down spin PDOS are denoted by thin solid line, thick solid line and dashed line, respectively.}
 \label{Fig8-NmMagTetra}
\end{figure}

In order to study the difference between magnetic and non-magnetic solutions in more detail, we superimpose the results of magnetic and non-magnetic calculations for Fe 3$d$ and As 4$p$ PDOS in Fig. \ref{Fig8-NmMagTetra} for tetragonal structure. Fe 3$d_{z^2}$ spin-orbitals exhibit two features separated by about 2 eV in the non-magnetic phase. Fe 3$d_{x^2-y^2}$ exhibit similar structures with somewhat larger separation ($\sim$ 2.5 eV). Both these bands do not contribute at $\epsilon_F$. Fe 3$d_{xy}$ and (3$d_{xz}$ + 3$d_{yz}$) bands possess one intense feature in each case and appear in the vicinity of $\epsilon_F$ providing dominant contribution at $\epsilon_F$. In the magnetically ordered phase, each band splits into up and down spin bands due to the exchange interactions. Such splitting is found to be about 2 eV in 3$d_{z^2}$ and 3$d_{x^2 - y^2}$ bands, while it is about 2.5 eV for all other cases. All the up spin bands appear below $\epsilon_F$ and the down spin states having 3$d_{xy}$, 3$d_{xz}$ and 3$d_{yz}$ symmetries primarily contribute at the Fermi surface indicating an essentially half-metallic scenario. As 4$p$ states exhibit interesting scenario in magnetically ordered phase. In addition to a small enhancement of As 4$p_z$ contributions at $\epsilon_F$, these states move towards lower energies in the magnetically ordered phase similar to the scenario in Fe 3$d$ case and exhibit small exchange splitting induced by Fe 3$d$ moments. In sharp contrast, the contribution of As (4$p_x$ + 4$p_y$) at $\epsilon_F$ in the magnetic phase is almost vanished, the energy bands move in opposite direction (towards higher energies as shown in Fig. \ref{Fig8-NmMagTetra}(f)) and does not show exchange splitting. Absence of exchange splitting may be apprehended due to the absence of hybridization of these states with the Fe 3$d$ states (As layers are situated above and below the Fe layers). However, energy shift in opposite direction is curious.

\begin{figure}
 \centering
 \vspace{-2ex}
 \includegraphics[trim={0cm 0cm 0cm 0cm}, clip, scale=.5]{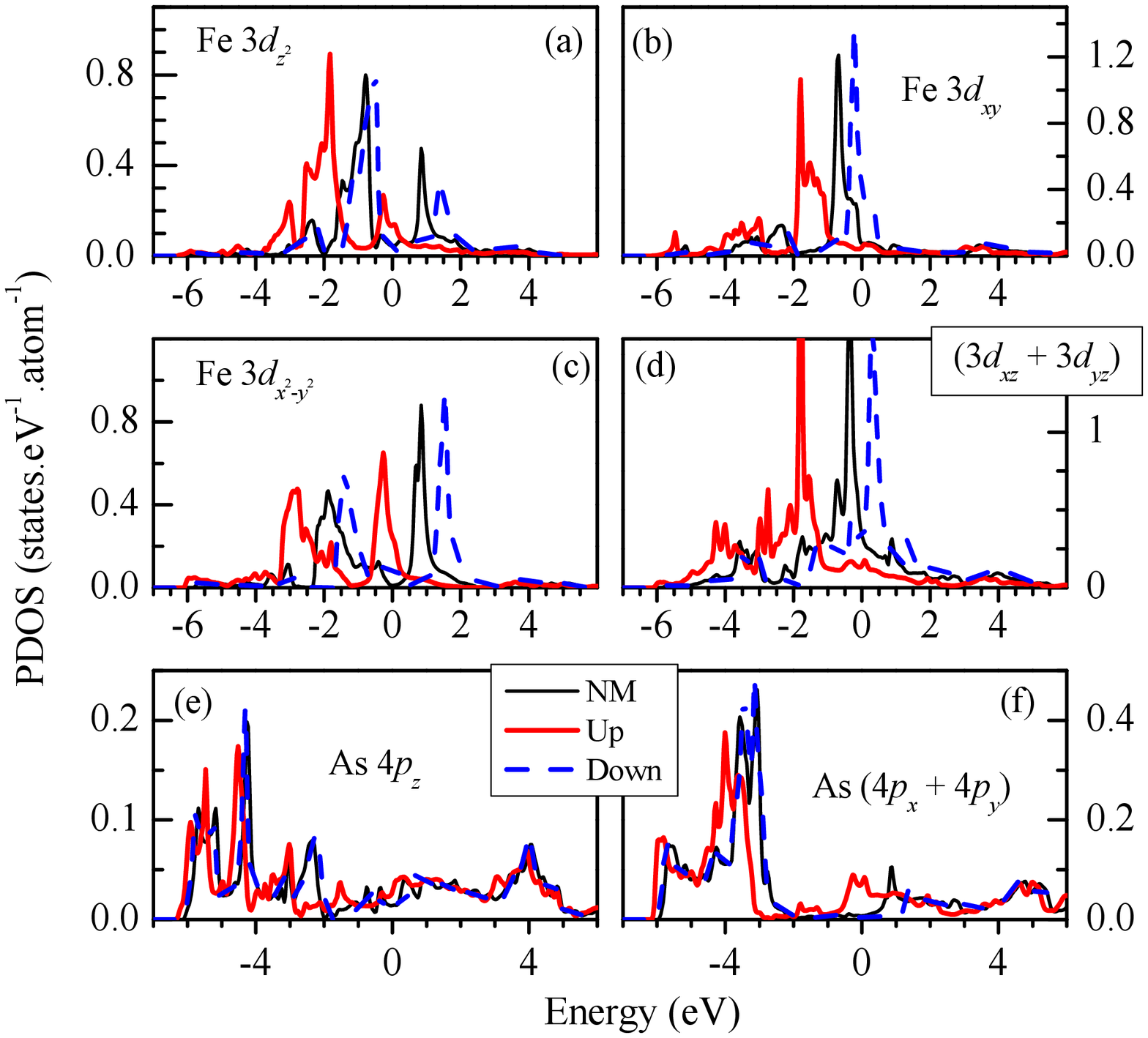}
    \vspace{-20 ex}
 \caption{Comparison of the partial Density of states of different Fe 3$d$ orbitals calculated for magnetic and non Magnetic (NM) configurations for cT$_{p2}$ structure. The details of the $d$ symmetries are defined as legends in each block of the plots. NM PDOS, and Up \& Down spin PDOS are denoted by thin solid line, thick solid line and dashed line, respectively.}
 \label{Fig9-NmMagcTp2}
\end{figure}

The scenario is somewhat different in cT phase - the results for cT$_{p2}$ phase is shown in Fig. \ref{Fig9-NmMagcTp2}. While the overall structure of the PDOS appear quite similar to the tetragonal case, there are significant differences. For examples, Fe 3$d_{z^2}$ and 3$d_{xy}$ PDOS exhibit exchange splitting of about 1 eV, which is much smaller than that in T phase. The behavior of 3$d_{x^2 - y^2}$ and 3$d_{xz}$ + 3$d_{yz}$ PDOS, however, appear similar to the tetragonal case. Interestingly, the contrasting behavior of As 4$p_z$ and (4$p_x$ + 4$p_y$) bands is not present in this case - $both$ the bands show small exchange splitting and an energy shift towards lower energies. A compression along the $c$ axis in the cT phase brings Fe and As atomic layers closer to each other and hence, the electronic states within As layer ($xy$-plane) also experiences Fe 3$d$ moment induced interactions. Here, the contribution of up spin (4$p_x$ + 4$p_y$) bands at $\epsilon_F$ becomes finite, while it was absent in the non-magnetic case. It is to note here that the down spin contributions corresponding to 3$d_{xy}$ and  3$d_{xz}$ + 3$d_{yz}$ states are essentially unoccupied with $\epsilon_F$ pinning at the bottom of the conduction band. However, these bands are significantly occupied in cT phase with $\epsilon_F$ appearing close to the peak of the bands.

\subsection{Fermi surface}

\begin{figure}
 \centering
 \includegraphics[scale=0.3]{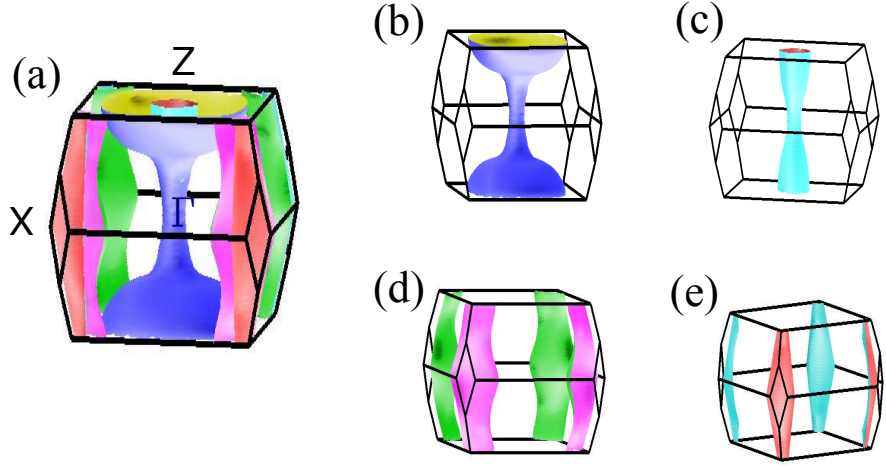}
 \caption{(a) Fermi surface in the tetragonal phase with relaxed $z_{As}$ ($\sim$~0.353); for better clarity in the demonstration of the Fermi sheets, the individual Fermi sheets are shown in (b) -(e).}
 \label{Fig10_tetra_rlx_fs}
 \end{figure}

\begin{figure}
 \centering
 \hspace{0ex}
 \includegraphics[scale=0.25]{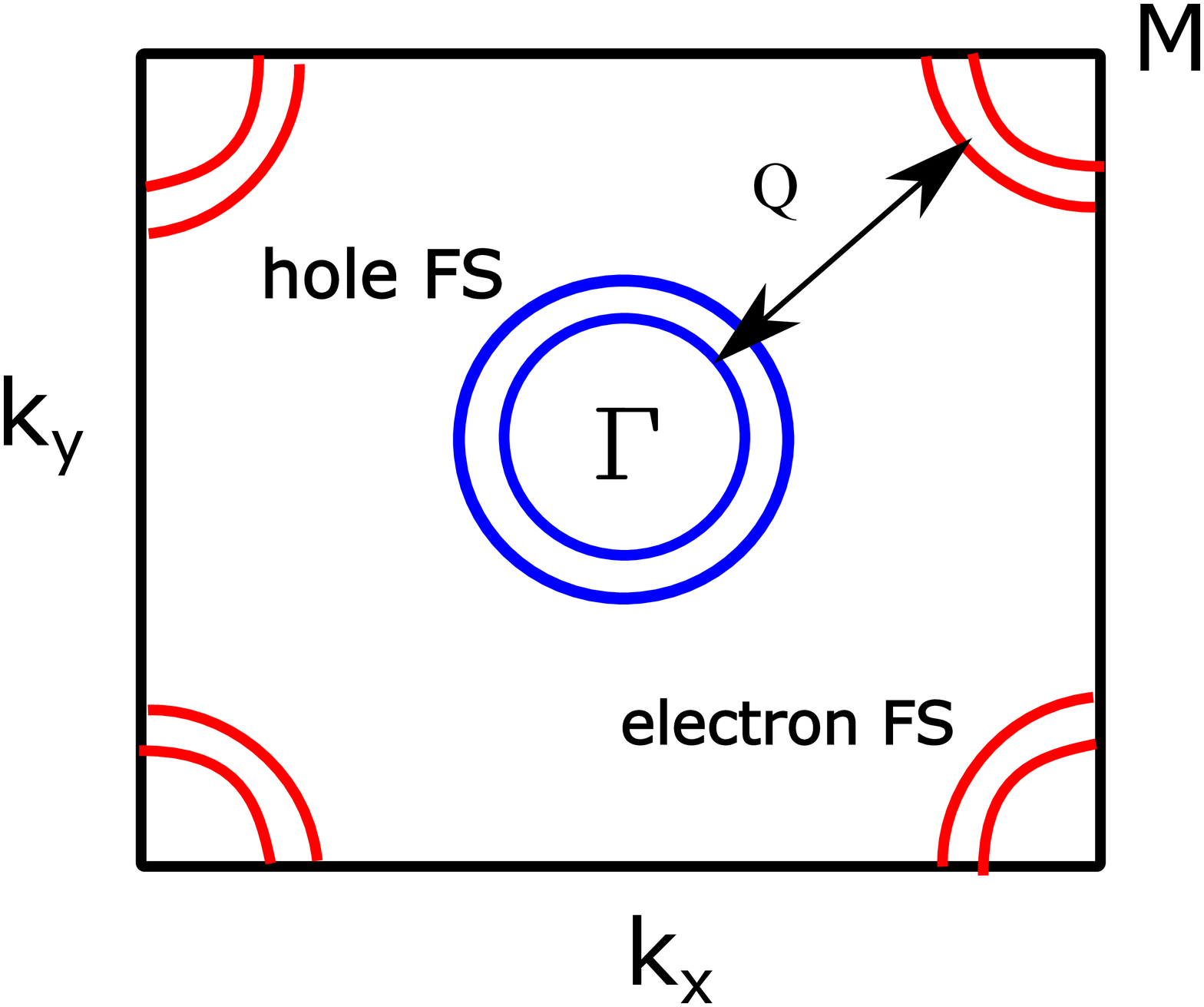}
 \caption{Schematic showing horizontal cut of Fermi surface in the tetragonal phase with relaxed $z_{As}$ (= 0.353). In the primitive tetragonal unit cell, the $M$ point corresponds to the $X$ point of body centered tetragonal unit cell.}
 \label{Fig11_fs_2d}
\end{figure}

In order to understand the Fermiology and its link to different crystal structures, the Fermi surfaces (FS) of CaFe$_2$As$_2$ are investigated below. Fig.\ref{Fig10_tetra_rlx_fs} depicts the Fermi surface in the tetragonal structure exhibiting quasi 2$D$ symmetry with the cylindrical axis as $k_z$-axis. In these calculations, $z_{As}$ is to optimize the structural parameters. It consists of four distinct Fermi sheets - two of them are wrapped around the $k_z$ axis at the brillouin zone center, $\Gamma$ and represent the hole states. Two Fermi sheets centered at the corner of the Brillouin zone are the electron pockets. The spin density wave type magnetic order observed in CaFe$_2$As$_2$ is associated to the Fermi surface nesting; a hole pocket centered around $\Gamma$-point is nested with an electron pocket around $M$-point. The dimensionality of the Fermi surface plays an important role in FS nesting; the 2$D$ behavior enhances the probability of nesting because parallel hole and electron sheets results in particle hole excitation and manifests as a Fermi surface nesting. The nesting wave vector $Q$ along $\Gamma - M$ direction in the primitive tetragonal unit cell is shown in Fig. \ref{Fig11_fs_2d}. Since, the nesting vector $Q$ is almost equal for both the hole sheet, both contribute in magnetic ordering.

\begin{figure}
 \centering
 \includegraphics[scale=0.25]{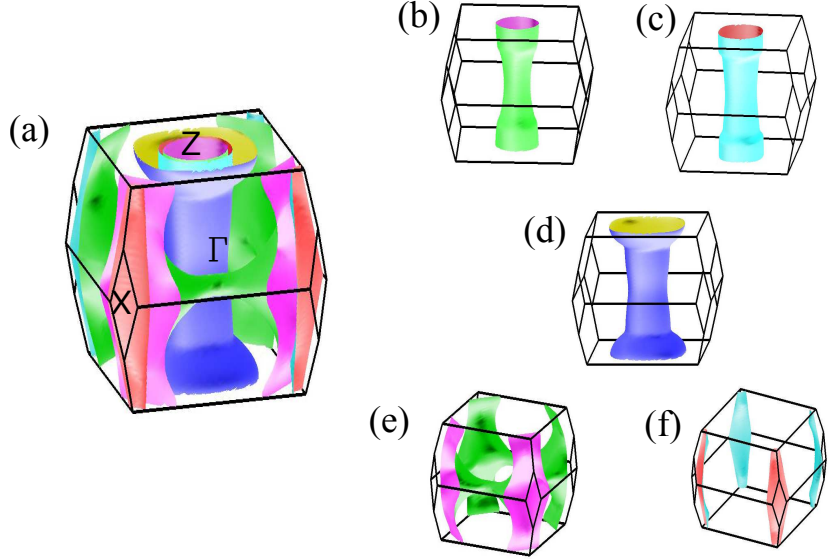}
 \caption{(a) Fermi surface in the tetragonal phase with $z_{As}$ fixed to its experimentally observed value of 0.372. (b) -(f) shows the individual Fermi sheets.}
 \label{Fig12_tetra_fs}
\end{figure}

The dimensionality and the topology of the Fermi surface depends on $z_{As}$ that is linked to the {\it pnictogen height}. The Fermi surface calculated with the experimental parameters (without employing relaxation in the structure) is shown in Fig. \ref{Fig12_tetra_fs}. In this case, the Fermi surface consists of five sheets; the additional hole sheet appears at the zone center due to lifting of the degeneracy of one of the hole Fermi sheets in the relaxed structure. One of the electron sheet spreads around the perimeter of the Brillouin zone, the other remains unchanged. The change in the shape of the Fermi sheets at the zone corners and the zone center leads to the disappearance of the nesting in one of the hole sheets.

From various calculations, we observe that the electronic structure corresponding to the experimentally found crystal structure is significantly different from the one obtained for the optimised structure within the model calculation via relaxing As positions. The relaxed structure favors smaller $z_{As}$ value and hence, smaller {\it pnictogen height} compared to the experimentally observed values and larger degree of Fermi surface nesting, thereby enhancing the degree of magnetic order. The real material, however, forms with relatively larger pnictogen height and one nesting vector as also observed in ARPES measurements of various materials in similar class. One reason for such difference could be related to the underestimation of correlation induced effects in the density functional calculations - the system prefers to reduce the hybridization and hence itineracy via increasing the pnictogen height. The other possibility could be a proximity of the ground state crystal structure of CaFe$_2$As$_2$ to a metastable state and thus, the system gets arrested to different strained crystal structure during sample preparation instead of going to the ground state configuration. The finding of the change of volume fraction of different structures and stains in SrFe$_2$As$_2$ with preparation conditions \cite{Saha-PRL-SrFe2As2} could be a realization of this conjecture. The experimental findings of contrasting behavior in CaFe$_2$As$_2$ is also in line with this view.

\begin{figure}
\centering
\includegraphics[scale=0.4]{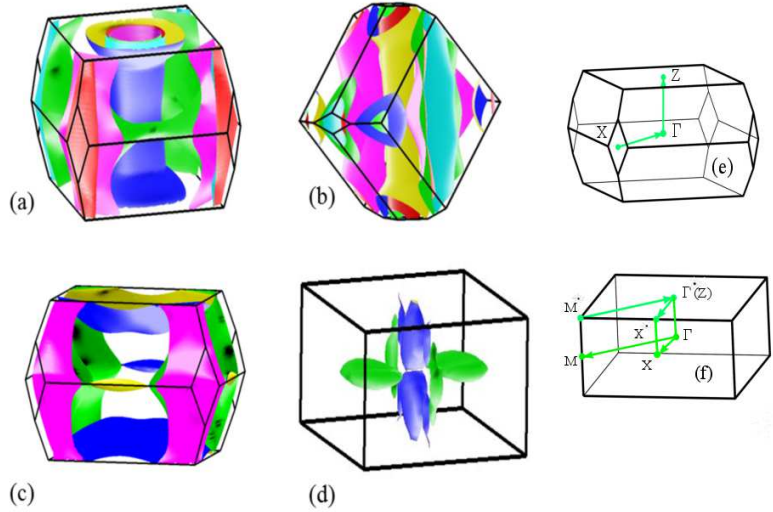}
\caption{ Fermi surfaces of CaFe$_2$As$_2$ in (a) tetragonal, (b) orthorhombic (c) collapsed tetragonal (cT phase at $P$ = 0.67 GPa) and (d) Antiferromagnetic orthorhombic structures. (e) and (f) show the Brillouin zone of the body centered tetragonal and tetragonal structures, respectively.}
\label{Fig13_fs}
\end{figure}

In Fig. \ref{Fig13_fs}, we investigate the Fermi surfaces in all the structural phases realized in CaFe$_2$As$_2$ by varying parameters like temperature, pressure etc. In the O phase, without any magnetic order, the Fermi surface looks very similar to the T phase. The difference in the definition of $k$ vectors appears due to their Brillouin zone orientation. It is evident from the comparison of Figs. \ref{Fig13_fs}(a) and \ref{Fig13_fs}(b) that the Fermi sheets in the orthorhombic phase are shrinked compared to those in the tetragonal phase as evident from the larger separation between electron and hole sheets. The orthorhombic phase emerges at low temperature presumably due to thermal compression leading to a small distortion of the lattice in the $xy$ plane and small ($\sim$ 5\%) compression in the $c$ axis. Thus, the pnictogen height becomes smaller as reflected in the Fermiology.

Fermi sheets in the O phase with AFM ordering are shown in Fig.  \ref{Fig13_fs}(d). Here, the Fermi sheets look three dimensional and very different from all other cases due to the opening of energy gap at $\epsilon_F$ induced by the nesting of the electron sheets at the zone corner with the hole sheets at zone center.

Notable difference appears in the cT structure; the Fermi surface shown here correspond to the structural parameters at $P$ = 0.67 GPa with $z_{As}$ value slightly smaller than the case shown in Fig. \ref{Fig10_tetra_rlx_fs}. We chose these parameters corresponding to slightly higher pressure than the previous cases to demonstrate the changes in Fermi surface topology with greater clarity. The 2$D$ nature of the hole Fermi surfaces around the $z$-axis is lost completely in the results for the cT phase. Here, all the three hole pockets around the $\Gamma$-point disappeared. Only one hole sheet survives around the $Z$ point making it a closed one with larger radius. The electron sheets are not affected significantly. Such drastic change in the hole Fermi surface around $k_z$-axis rules out the electron-hole nesting responsible for magnetic long range order and is consistent with the experimental finding of absence of magnetic order in this phase.

\subsection{Energy bands}

\begin{figure}
\centering
 \includegraphics[scale=0.25]{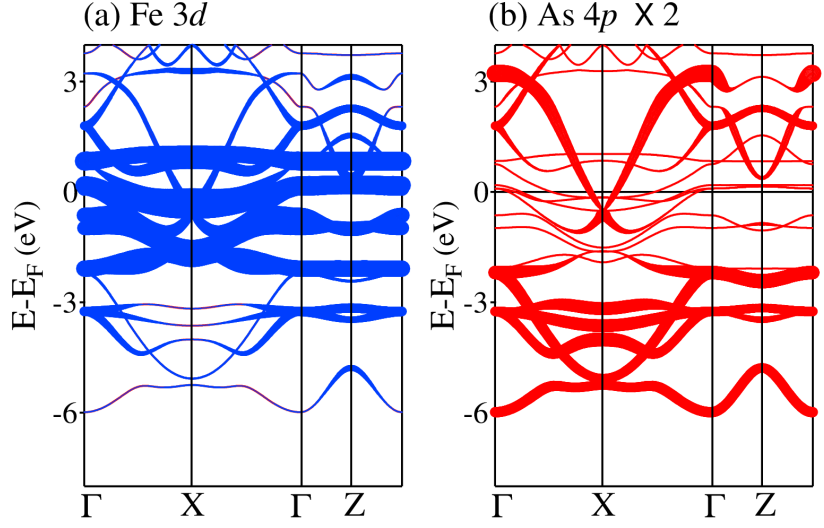}
\caption{Band structure of CaFe$_2$As$_2$ in the tetragonal structure highlighting (a) Fe 3$d$ and (b) As 4$p$ character; the thickness of lines represents the weight of the character. As 4$p$ contributions are rescaled by a factor of 2 to enhance the visibility of the contributions.}
\label{Fig14_bonding}
\end{figure}

\begin{figure}
\centering
\includegraphics[trim={1cm 0cm 0cm 0cm}, clip, scale=0.25]{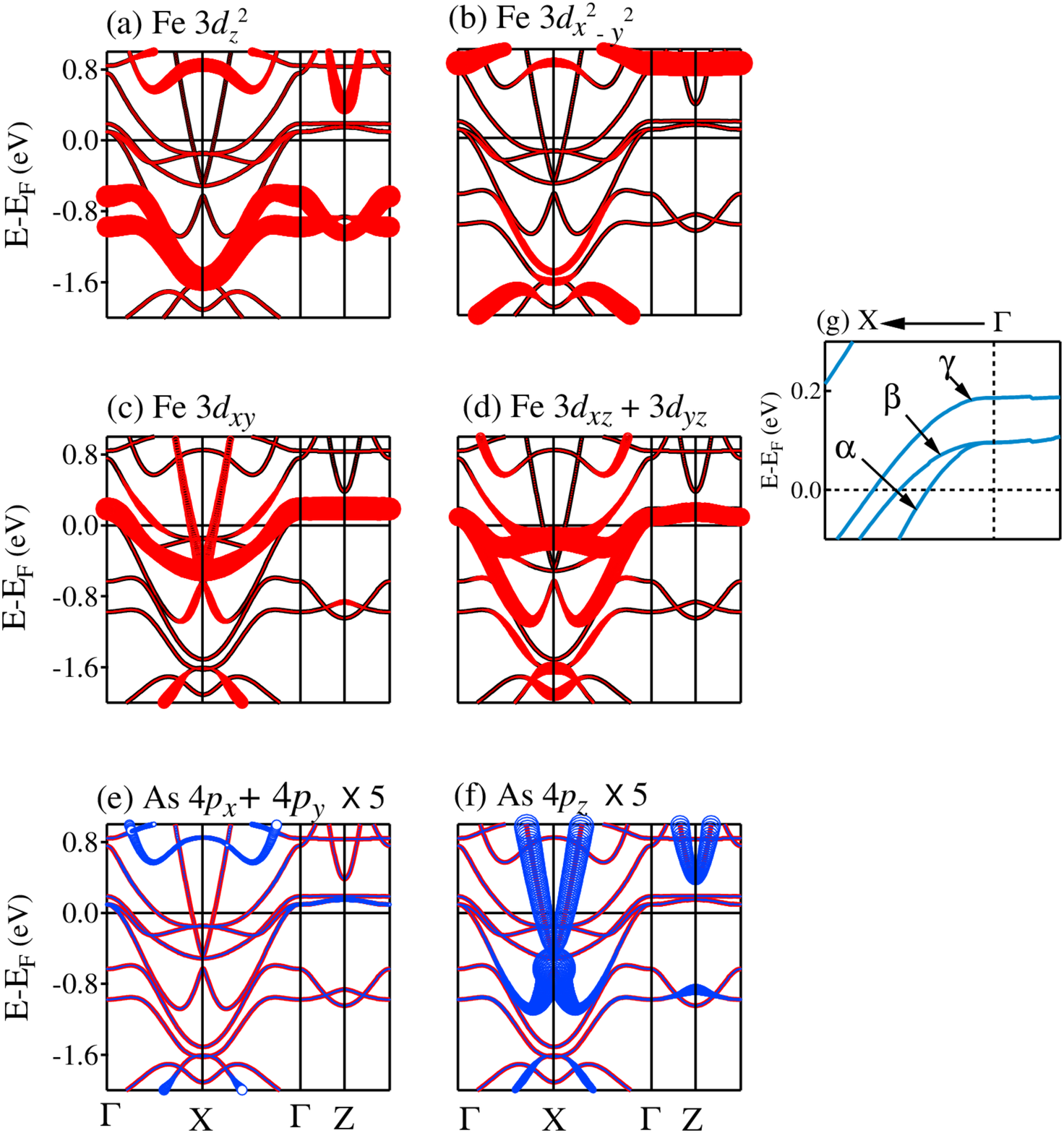}
\caption{Energy bands of tetragonal CaFe$_2$As$_2$ near the Fermi level. The thickness of the lines exhibit (a) Fe 3$d_{z^2}$, (b) Fe 3$d_{x^2-y^2}$ , (c) Fe 3$d_{xy}$, (d) Fe (3$d_{xz}$ + 3$d_{yz}$), (e)  As (4$p_x$ + 4$p_y$), (f) As 4$p_z$ characters. (g) Energy bands along $\Gamma-X$ direction in a highly expanded energy scale to distinguish them with clarity. The bands are defined as $\alpha$, $\beta$ and $\gamma$ bands.}
\label{Fig15_t_band_char}
\end{figure}

We now turn to the investigation of the properties of the energy bands. While the density of states discussed above provide information about the contribution of various electronic states as a function of energy, these results will help to understand the $k$-resolved information, which are often helpful to understand the electronic properties and can also be observed in angle resolved photoemission measurements. In Fig. \ref{Fig14_bonding}, we show the energy bands of CaFe$_2$As$_2$ for the tetragonal structure; the thickness of the line represents the orbital contributions. It is evident that the Fe 3$d$ symmetry adapted bonding bands appearing between -6 to -2 eV possess dominant As 4$p$ character. The bands close to the Fermi level are primarily constituted by Fe 3$d$ states with notable contributions from As 4$p$ states. Interestingly, while the bands around $\Gamma$ point forming the hole pocket in the Fermi surface possess essentially Fe 3$d$ character, one of the electron pocket around $X$-point possess large As 4$p$ character.

The detailed orbital symmetries near $\epsilon_F$ are shown in Fig. \ref{Fig15_t_band_char}. At $\Gamma$, three energy bands are observed to cross $\epsilon_F$ - these bands are denoted by $\alpha$, $\beta$ and $\gamma$ as shown separately in Fig. \ref{Fig15_t_band_char}(g) for better clarity. These three energy bands create three hole pocket around $\Gamma$ \cite{ARPES-Liu}. There are two electron pockets around the high symmetry point, $X$. It is clear from Fig. \ref{Fig15_t_band_char}(c) and Fig. \ref{Fig15_t_band_char}(d) that the $\gamma$ band possesses mainly $d_{xy}$ character and the $\alpha$ \& $\beta$ bands have contribution from degenerate $d_{xz}$ and $d_{yz}$ orbitals \cite{ganesh-CaFe2As2}. These three bands possess $t_{2g}$ symmetry. The $e_g$ bands constituted by $d_{x^2-y^2}$ and $d_{z^2}$ spin-orbitals appear far away from the Fermi level and hence do not play significant role in the electronic properties of the system. Overall, the bands near the Fermi level has very less As 4$p$ character except one electron pocket at $X$, which has As 4$p_z$ character. This is in line with the expectation from the fact that As layers appear above and below Fe layers and hence $p_z$ orbital is expected to play a major role in the Fe 3$d$-As 4$p$ hybridization.

\begin{figure}
 \centering
 \includegraphics[scale=0.5]{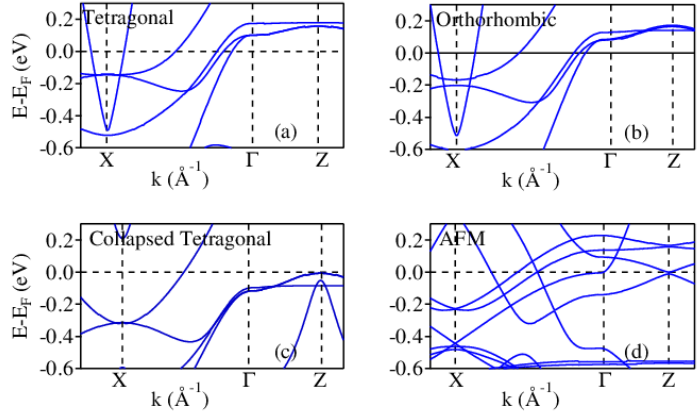}
 \caption{Calculated energy bands of (a) tetragonal phase, (b) orthorhombic phase, (c) non magnetic collapsed tetragonal phase and (d) antiferromagnetically ordered orthorhombic phase}
 \label{Fig16_band_xgz}
 \vspace{-2ex}
\end{figure}

In Fig. \ref{Fig16_band_xgz}, we show the change in band structure in different phases along $X - \Gamma - Z$ symmetry lines. In collapsed tetragonal phase, the bands at $\Gamma$ shifts below the Fermi level and the hole pockets around $\Gamma$ disappear. The nonmagnetic calculation of the orthorhombic structure gives rise to similar band structure like tetragonal phase, but if we consider the antiferromagnetic configuration then the band structure becomes complex. The hole pocket at $\Gamma$ nesting with an electron pocket at $X$ gives rise to an energy gap at $\epsilon_F$ via folding of the energy bands at the zone boundary created by the magnetic supercell.

\section{Discussion}

From the above results, it is clear that electronic structure of CaFe$_2$AS$_2$ is very sensitive to the crystal structure and {\it pnictogen height} is an important parameter in deriving the ground state properties. The system appears to favor smaller {\it pnictogen height} although the experiments found structures with somewhat higher values and significant dependence of the structure on preparation conditions. On application of pressure, the tetragonal symmetry gets arrested till the low temperatures and the electronic properties become significantly different from magnetically ordered phase observed in ambient conditions.

In cT phase, the $c$-axis collapse reduces Fe moment and Fe-As hybridization along $z$ direction enhances \cite{dft_yildrim_prl2009}. In fact, the change in Fe-As hybridization and consequent change in the magnetic moment of Fe turn out to be the key factors in deciding the magnetic and structural properties of CaFe$_2$As$_2$. To get a clear idea about this we have calculated the local magnetic moment at Fe-sites. The calculation are done for tetragonal phase and collapsed tetragonal phase. In the later case, we choose the lattice parameters corresponding to two different applied pressures. The phase realized under 0.47 GPa pressure is termed as cT$_{p1}$ and the cT phase with $P$ = 0.63 GPa is called cT$_{p2}$. Experimentally, the magnetization in the cT phase has been studied extensively and no signature of long range magnetism was found. There are controversy on the magnetic moment though - while Ma Long {\it et al.} \cite{MaLong-ChinesePhysB13} found signature of quenching of Fe moment on transition to cT phase induced by Pr substitution at Ca sites, magnetic susceptibility data by W. Wu {\it et al.} \cite{ct_sc_absence_Yu_PRB_2009} indicate finite moment of Fe in the cT phase though depleted from its value in ambient tetragonal phase. Moreover, the Fe magnetic moment appears to be sample dependent. We calculated Fe moments assuming parallel alignment of Fe-moments and compare the results with their values corresponding to the non magnetic phase in order to investigate the trend of the change of converged energies with the change in external parameters, which is often captured well employing DFT methodologies \cite{Andersen-PRL,CaSrRuO-PRB,BaIrO3-PRB}. The table below gives the list of lattice parameters used for the calculations along with the calculated magnetic moment and ground state energies.

\begin{table}
\caption {The magnetic moment of Fe and ground state energy at various structural configuration of CaFe$_2$As$_2$, the experimental latice parameters are taken from \cite{neutron-kreyssing-PRB08}}
\label{tab:table1}
\begin{center}
\begin{tabular}{ | c | c | c | c | c | c | c | c | c | c |}
\hline\hline
Phase & $p$ & $a$ & $c$ & $z_{As}$ & $d_{FeAs}$ & $d_{AsAs}$ & $\mu_{Fe}$  & $E(Mag)$ & $E(NM)$ \\

& (GPa) & (\AA) & (\AA) &  &  (\AA) &  (\AA) & ($\mu_B$) & (meV) & (meV) \\
\hline
T \hfill\ & 0 & 3.8915 & 11.690 & 0.372 & 2.4145 & 2.9926 & 2.37 & 0 & 150 \\

\hline
cT$_{p1}$  & 0.47 & 3.9785 & 10.6178 & 0.3657  & 2.3380 & 2.8519 & 1.37 & 60 & -198 \\

\hline
cT$_{p2}$ \hfill  & 0.63 & 3.9780 & 10.6073 &0.3663  & 2.3405 & 2.8364 & 1.95 & 100 & -194 \\

\hline\hline
\end{tabular}
\end{center}
\end{table}

In the tetragonal structure, the converged energy for the magnetically ordered configuration is about 150 meV lower than that for the nonmagnetic (NM) configuration suggesting a preference for the magnetic order in the ground state. The magnetic moment of Fe atoms is found to be 2.37~$\mu_B$ in this structure. The cT structure manifests a contrasting scenario exhibiting significantly higher energy for magnetically ordered phase relative to the NM phase. Application of pressure leads to a gradual reduction of the lattice constant, $c$ and As-As bondlength ($d_{AsAs}$). The magnitude of the lattice constant, $a$ and $z_{As}$ exhibit anomalous scenario with the largest and smallest values at $P$ = 0.47 GPa, respectively. With the application of pressure, the nonmagnetic phase appears to be favored more. The ground state energy of the non-magnetic phase at cT$_{p1}$ structure is found to be the lowest relative to all the calculations we have done.

The Fe moment appears to diminish with the application of pressure. Interestingly, the lowest moment of 1.95~$\mu_B$ is found for the pressure of 0.47 GPa (cT$_{p1}$ phase) and then the Fe moment increases with further increase in pressure although the compression of $c$ is more at cT$_{p2}$ phase compared to cT$_{p1}$ phase. The value of $z_{As}$, however, is smaller in the cT$_{p1}$ phase indicating a direct link of the Fe-moment to $z_{As}$ value. It is still an open question whether the magnetic moment at Fe sites actually vanishes or there is a small magnetic moment on Fe atom, but they don't give any long range order \cite{dft_yildrim_prl2009}. All our calculations indicate that the magnetic configuration in the cT structure has higher energy compared to the non-magnetic phase. The magnetic moment decreases as the Fe-As bond length decreases. This can be understood considering the fact that decrease in Fe-As bondlength makes the conduction electrons more itinerant and hence less local moment. It is puzzling that the Fe-As bondlength is smallest in cT$_{p1}$ structure and a small decrease in Fe-As bond length from 2.3405 \AA\ in cT$_{p2}$ structure to 2.338 \AA\ in cT$_{p1}$ structure changes the magnetic moment from $1.95 \mu_B$ to  $1.37 \mu_B$.

The above results establish that the Fe spin state is highly sensitive to the Fe-As hybridization, which is linked to the {\it pnictogen height} too. The shorter As-As distance leads to increased hybridization within the As-plane and higher degree of As $p$ electron itineracy. However, the local character of the Fe $d$ electrons remain less affected. On the other hand, higher Fe-As hybridization leads to larger covalent splitting of the Fe-As bonding and antibonding bands leading to a shift of the valence bands. Some of the bands shifts below $\epsilon_F$ and become completely filled. The energy bands crossing the Fermi level will have higher itineracy and hence less local character, which is presumably one of the reasons for the reduction of Fe moment and the resistivity.

The calculated Fe-As bond length in the AFM ground state with orthorhombic structure is 2.3654 \AA, which is larger than that in cT phase and shorter than the one in T phase. The Fe spin moment is found to be 1.87 $\mu_B$, which is much larger than the experimental value of 0.80 $\mu_B$. Various theoretical studies tried to capture the experimental moment by varying $z_{As}$ (Experimental value = 0.3664). It is observed that the spin moment becomes closer to the experimental value for the $z_{As}$ values of 0.3567 \cite{structure_Tompsett_physicab2009}. It is clear that $z_{As}$ and hence the {\it pnictogen height} is an important parameter leading to varied electronic properties in this system. Moreover, these results establish that the cT phase possesses finite magnetic moment as also found experimentally in magnetic susceptibility measurements for almost all kinds of phases prepared so far. The absence of long range order appears to be due to its proximity to the critical point of loosing Fermi surface nesting \& no magnetic order \cite{ganesh-cr,cao-casrruo,maiti-casrruo}.

\begin{figure}
 \centering
 \includegraphics[scale=0.5]{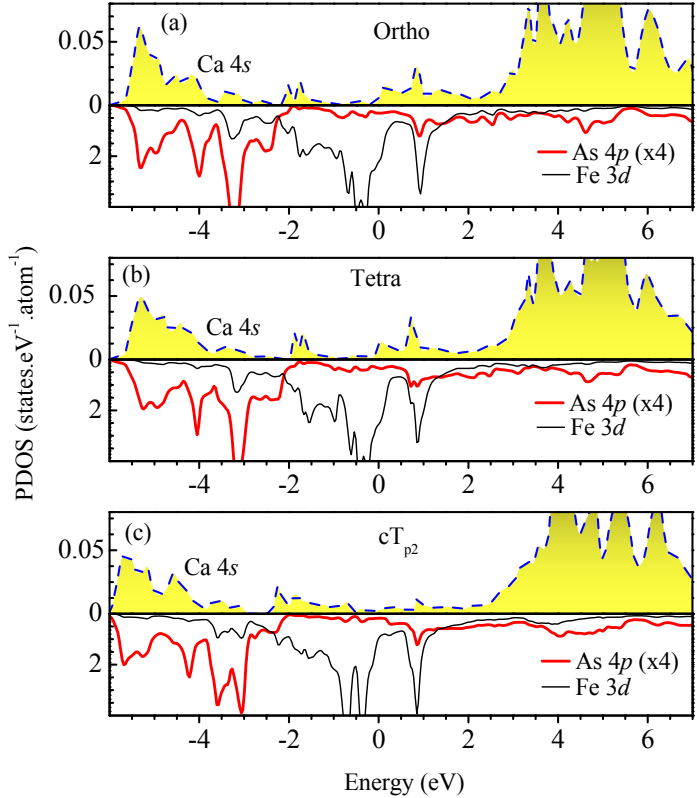}
\vspace{-8ex}
 \caption{Ca 4$s$ PDOS (area plot) is compared with As 4$p$ (thick solid line) and Fe 3$d$ thin solid line) for (a) orthorhombic, (b) tetragonal and (c) collapsed tetragonal structures. The As 4$p$ contributions are rescaled by 4 times for better clarity.}
 \label{Fig17_chReserv}
 \vspace{-2ex}
\end{figure}

Interestingly, coupling of Ca 4$s$ states to the conduction layer becomes stronger in the cT phase as discussed in Fig. \ref{Fig6TcT-As4p} and redrawn here to emphasize the scenario. In Fig. \ref{Fig17_chReserv}, we show the Ca 4$s$ PDOS along with Fe 3$d$ and As 4$p$ contributions for different structures. Ca 4$s$ contributions in orthorhombic and tetragonal phases look almost similar with major intensities above $\epsilon_F$, a reasonably intense feature between -6 to -3 eV and almost no intensity in the vicinity of $\epsilon_F$. While the intensity above $\epsilon_F$ remains dominant in the cT phase too like other cases, there is significant redistribution of PDOS in the lower energy regime. Most importantly the contribution in the vicinity of $\epsilon_F$ becomes almost double of the contributions in other two structures. The contributions in the energy range -3 to -6 eV appear due to the Fe-As bonding bands with dominant As 4$p$ character and -2.5 to 0 eV is contributed by anti-bonding bands with dominant Fe 3$d$ character. The intense feature between -2.5 to -4 eV in As 4$p$ PDOS are the non-bonding As 4$p$ states. The features in Ca 4$s$ PDOS resemble well with structures in the As 4$p$ PDOS as these states emerges due to the hybridization of Ca 4$s$ states with As 4$p$ states. Thus, the enhancement of 4$s$ contributions below $\epsilon_F$ can be attributed to enhanced hybridization of Ca and As states due to compression of $c$ axis in the cT phase.

Charge reservoir layers, generally insulating in nature, are believed to play an important role in superconductivity \cite{charge-reservoir,charge-reservoir_book} as these layers help to preserve two dimensionality of the system and can protect the conduction layers via screening of substitutional and/or other type of disorder induced effects. Effective two dimensional electronic structure seem to be essential to achieve high superconducting transition temperature and almost all the superconductors (conventional and unconventional) having high transition temperature possess two dimensional electronic structure. Thus, the higher degree of Ca-As hybridization in collapsed tetragonal phase and the observation of pressure induced superconductivity \cite{edoped_sc_cafe2as2_Lv_pnas_2011,sc_cafe2as2_Jeffries_PRB_2012,sc_cafe2as2_Neeraj_PRB_2009,NMR_kawasaki_iop}
appears to be linked.

\section{Conclusion}

In summary, we present here the detailed electronic structure of CaFe$_2$As$_2$ for different crystallographic structures using density functional theory. The calculated density of states indicate strong covalency induced effect in the electronic structure, which increases with the compression of $c$ axis observed in collapsed tetragonal phase. While the energy distribution of density of states appear similar in tetragonal and orthorhombic phase in ambient conditions, a significant shift towards lower energies is observed in collapsed tetragonal phase, which reduces the total energy of the system. The hole pockets at the $\Gamma$ point eventually vanishes at higher pressure ruling out the Fermi surface nesting related to the formation of spin density wave states. The electron pockets centered around the high symmetry points $M$/$X$/$Z$ evolves significantly in the cT phase. The enhancement of Fe-As hybridization in the cT phase presumably leads to higher degree of itineracy relevant for the changes in electrical resistivity in this system. Interestingly, the changes in Ca 4$s$ partial density of states suggest enhancement of Ca-As hybridization with the decrease in $c$, which is significant in the context of superconductivity as it is believed that such layers act as charge reservoir layer in various high temperature superconductors. We observe that the changes in the Fermiology and energy band shifts due to application of pressure depletes the magnetic moment at Fe-sites although they do not show signature of magnetic order indicating the proximity of such strained condition to the quantum fluctuations.

\section{Method}

Electronic structure calculations were carried out using full potential linearized augmented plane wave method (FLAPW) captured in the Wien2k software \cite{wien2k}. Convergence was achieved by fixing the energy convergence criteria to 0.0001 Rydberg (1 meV). For nonmagnetic calculations, we have used 10$\times$10$\times$10 $k$-points in the Brillouin zone and for Fermi surface calculations 39$\times$39$\times$10 $k$-points were used. We have used the Perdew-Burke-Ernzerhof generalized gradient approximation (GGA)\cite{dft_gga} for our density functional theoretical calculation \cite{dft_ks}. The Fermi surfaces were calculated using Xcrysden\cite{crysden}. The collapsed tetragonal phase of CaFe$_2$As$_2$ achieved via application of pressure/varying annealing conditions, possesses the same space group but a reduced $c$ axis and slightly increased $a$ axis. The volume of the unit cell reduces effectively. The lattice parameters at $P$ = 0.63 GPa are $a$ = 3.9780(1)\AA, $c$ = 10.6073(7)\AA, and $z_{As}$ = 0.372(1)\cite{neutron-kreyssing-PRB08}.
Orthorhombic structure of CaFe$_2$As$_2$ appearing at low temperature and ambient pressure has $Fmmm$ space group with the lattice parameters $a$ = 5.506(2)\AA, $b$ = 5.450(2)\AA, $c$ = 11.664(6)\AA, and $z_{As}$ = 0.36642(5)\cite{neutron-kreyssing-PRB08}. For Antiferromagnetic calculations, we used 10$\times$10$\times$10 $k$-points in the Brillouin zone and for Fermi surface 23$\times$23$\times$10 $k$-points.





%

\section*{Acknowledgements}

K. M. acknowledges financial assistance from the Department of Science and Technology,
Govt. of India under J.C. Bose Fellowship program and Department of Atomic Energy, Govt. of India.

\section*{Author Contributions}

K.A. carried out all the calculations. K.M. initiated the study and
supervised the project. K.A. and K.M. jointly analyzed the data
and prepared the manuscript.

\section*{COMPETING FINANCIAL INTERESTS}

The authors declare no competing financial interests.

\end{document}